\documentclass[twocolumn,showpacs,pra,aps,superscriptaddress,floatfix]{revtex4}
\usepackage{graphicx}

\begin{document}

\title{Quantum shutter approach to tunneling time scales with wave packets}

\author{Norifumi Yamada}
\email[]{yamada@i1nws1.fuis.fukui-u.ac.jp}
\affiliation{Department of Information Science, Fukui University\\
3-9-1 Bunkyo, Fukui, Fukui 910-8507, Japan}
\author{Gast\'on Garc\'{\i}a-Calder\'on}
\email[]{gaston@fisica.unam.mx}
\affiliation{Instituto de F\'{\i}sica,
Universidad Nacional Aut\'onoma de M\'exico, 
Apartado Postal {20 364}, 01000 M\'exico, Distrito Federal, M\'exico}
\author{Jorge Villavicencio}
\email[]{villavics@uabc.mx}
\affiliation{Facultad de Ciencias, Universidad Aut\'onoma de Baja California, 
Apartado Postal 1880, 22800 Ensenada, Baja California, M\'exico}

\date{\today}

\begin{abstract}
The quantum shutter approach to tunneling time scales (G. Garc\'{\i }a-Calder\'{o}n and 
A. Rubio, Phys. Rev. A \textbf{55}, 3361 (1997)), which uses a cutoff plane wave as the 
initial condition, is extended in such a way that a certain type of wave packet can be 
used as the initial condition. An analytical expression for the time evolved wave function 
is derived. The time-domain resonance, the peaked structure of the probability density 
(as the function of time) at the exit of the barrier, originally found with the cutoff 
plane wave initial condition, is studied with the wave packet initial conditions. It is 
found that the time-domain resonance is not very sensitive to the width of the packet 
when the transmission process is in the tunneling regime. 
\end{abstract}


\maketitle

\section{Introduction}

Tunneling is one of the most important quantum phenomena that has been widely applied 
in science and technology. For years, the stationary treatments of tunneling were 
sufficient for many practical purposes, and the details of tunneling dynamics 
were not urgent issues to investigate. This is not the case anymore. The interest in 
tunneling dynamics is increasing as, for example, the number of carriers involved in 
tunneling events decreases due to the rapid downsizing of semiconductor devices. In 
principle, the tunneling dynamics can be completely understood if one can solve the 
time dependent Schr\"odinger equation, taking other degrees of freedom into account 
that affect the tunneling particles. This is, however, a difficult task in general. 
It is thus important to study the time scales of tunneling dynamics (tunneling times) 
in simplified models and use them for qualitative understanding of the tunneling 
dynamics in realistic systems. There are many approaches to define or measure the 
tunneling times in simplified models \cite{HS89, LM94, review4, recami}, and the 
quantum shutter approach \cite{quantumshutter} is one of them. The present paper 
concerns a generalization of the quantum shutter approach. 

Let us consider a one dimensional scattering problem where a particle is incident on a 
potential $V(x)$. The time scales that characterize the tunneling dynamics are 
^^ ^^ embedded" in the wave function. To extract them from the wave function, the 
quantum shutter approach \cite{quantumshutter} uses a cutoff plane wave as the initial 
condition and monitors how the probability density changes in time at a specified position 
(e.g., at the exit of the barrier) or in a spatial region (e.g., in the well region in 
a double barrier structure) to find out the time scales that characterize the transient 
behavior of the wave function (from nonstationary to stationary). The use of a cutoff 
plane wave initial condition can be understood as an analogy to the use of a step input 
in the study of the temporal response of an electrical (e.g., RCL) circuit. 

Studies of tunneling dynamics involving cutoff plane waves go back back to Stevens \cite{stevens}, 
who argued the signal velocity under the barrier by applying the contour deformation technique 
developed by Brillouin \cite{brillouin}. This technique allows one to decompose a wave propagating 
in a dispersive medium into three parts: the fore-runners, the monochromatic part oscillating 
with the same frequency as the source, and the after-runners; the signal velocity is then defined 
as the velocity with which the monochromatic front moves. The under-the-barrier signal velocity 
found by Stevens was, however, questioned later by Teranishi \textit{et al} \cite{teranishi}, Jauho 
and Jonson \cite{jj89}, and Ranfagni \textit{et al} \cite{rma91}; these authors showed 
that the monochromatic front to which the signal velocity is attributed is in fact not 
appreciable in magnitude. Through these works, it was recognized that it is not appropriate 
to focus only on the monochromatic part of the wave. B\"uttiker and Thomas \cite{bt98} and 
Muga and B\"uttiker \cite{mb2000} thus studied not only the monochromatic part but also 
the fore-runners in detail, whereas Brouard and Muga \cite{bm96} turned their attention 
to the total wave function (under a cutoff plane wave initial condition) and studied its 
properties with an exact analytical expression for the wave function which they derived. 
Independently, Garc\'{\i }a-Calder\'{o}n and Rubio \cite{quantumshutter} derived an exact 
analytical expression for the wave function with a cutoff plane wave initial condition and 
applied it to the analysis of the transient behaviors in the tunneling dynamics. It may be 
said that these studies formed a new area of research in the field of quantum dynamics, 
where one explores the tunneling dynamics, especially its transient behaviors, by using 
exact analytical expressions for the wave functions with cutoff plane wave initial conditions. 
The quantum shutter approach \cite{quantumshutter} and the approach by Brouard and Muga \cite{bm96} 
use the same tool called $M$ function [see Eq. (\ref{eq:3})] to express the wave functions  
in analytical manners. They differ, however, in the following respect: in addition to the $M$ 
function, the quantum shutter approach uses the resonant eigenfunctions, while the approach 
by Brouard and Muga uses an entire function that arises from a pole expansion of a function 
involved in the momentum eigenfunction expansion of the wave functions. In \cite{equivalence}, 
an analytical expression for the transmitted wave was obtained in sole terms of the $M$ 
function and the poles and the residues of the transmission amplitude. Another interesting 
and unexpected aspect of the quantum shutter approach is that it has a close relationship 
with the consistent history approach to the tunneling time problem as shown in \cite{equivalence}. 
In particular, the probability density at the exit of a rectangular barrier under a cutoff plane 
wave initial condition, a quantity of major concern in the quantum shutter approach, coincides, 
when properly normalized, with a function $G_{\rm p}(t)$ that is defined in the consistent 
history approach. Function $G_{\rm p}(t)$ allows us to associate the transient behaviors of 
the wave function with the interference between Feynman histories with different tunneling 
times, which provides a novel viewpoint to the transient behavior. In this way, the quantum shutter 
approach is related not only to the stream of research that began with the work of Stevens 
but also to a relatively new approach, the consistent history approach, to the tunneling time 
problem. In Ref. \cite{yamadaprl}, the consistent history approach was used to give different 
tunneling times in a unified manner. 

The quantum shutter approach, formulated for a general (but finite-range) potential, 
has been applied to single barriers \cite{single1,single3,single4}, double barriers \cite{quantumshutter}, 
and to a more general superlattice structure \cite{superlattice}. For the case of 
tunneling through a single barrier, the probability density at the exit of the barrier 
was found to have a peaked structure, which is called the \textit{time-domain 
resonance}. The time $t_{\rm p}$, corresponding to the  peaked value of the resonance, 
has been studied in detail as a function of the system parameters and has been  proposed 
as a tunneling time scale \cite{single4}. For the case of tunneling through a multi-barrier 
resonant structure, the buildup process in the well region(s) has been also  studied, 
with the expectation that it has important implications on the speed of resonant tunneling 
devices \cite{quantumshutter, superlattice}. These 
studies, while demonstrating the usefulness of the quantum shutter approach, raise a 
natural question: To what extent do the results obtained from the approach, in particular 
the existence of the \textit{time-domain resonance}, depend on the special form of 
the initial conditions? To answer this question, it is necessary to extend the 
approach to a more general class of initial conditions. 

The purpose of the present paper is to provide  an extension of the quantum shutter approach 
to a type of wave packet initial condition, which reduces itself to the cutoff plane wave 
initial condition in the limit of an infinite packet width. This extension does not change 
the analytical scheme of the approach, so that the time evolved wave function continues to 
be available in an analytical form. We apply the extended approach to tunneling through a 
square barrier to study how the time-domain resonance is affected by the width of the initial 
wave packet. 

\section{Formalism}

The quantum shutter approach assumes an arbitrary but finite-range potential $V(x)$ [in 
this paper $V(x)$ is such that it vanishes for $x<0$ and for $x>d$] and uses, for example, 
the following form of the cut off plane wave as the initial wave function: 
\begin{eqnarray}
\Psi(x,0)=\cases{2i\sin k_0x & for $x<0$\cr 0 & for $x\ge 0$,}
\label{eq:1}
\end{eqnarray}
where $k_0$ is the wave number. This setup corresponds to the physical situation where a beam 
of particles with energy $E_0=\hbar^2k_0^2/2m$ (though not exactly monochromatic due to the 
sharp front of the wave) impinges on a shutter placed at $x=0$, just at the left edge of the 
potential; the tunneling process begins with the instantaneous opening of the shutter at $t=0$, 
enabling the incoming wave to interact with the potential for $t>0$. When the initial condition 
is given by Eq. (\ref{eq:1}), the exact solution of the time dependent Schr\"odinger equation 
along the transmitted region $x\ge d$ is found to be \cite{single1,equivalence} 
\begin{eqnarray}
\Psi(x,t)=T(k_0)M(x,k_0;t)-T(-k_0)M(x,-k_0;t)\nonumber\\
-2k_0\sum\limits_{n=-\infty }^{\infty}\frac{r_n}{k_0^2-k_n^2}M(x,k_n;t), 
\label{eq:2}
\end{eqnarray}
where $k_n$ is the $n$-th pole of the transmission amplitude $T$ (poles lie in the lower-half 
of the complex $k$ plane), $r_n$ is the associated residue, and the $M$ functions are defined by
\begin{eqnarray}
M(x,q;t)&\equiv&\frac{i}{2\pi}\int_{-\infty}^\infty\! dk\, 
\frac{e^{ikx-i\hbar k^2 t/2m}}{k-q}\label{eq:3}\\[2mm]
&=&\frac{1}{2}e^{(imx^2/2\hbar t)}w(iy_q), 
\label{eq:4}
\end{eqnarray}
where $q=k_n, \pm k_0$; the function $w(z)$, which is often called Faddeeva function, is related 
to the complex complementary error function as $w(z)=e^{-z^2}\textrm{erfc}(-iz)$ \cite{wfunction}, 
and $y_q$ is given by
\begin{equation}
y_q=e^{-i\pi /4}\sqrt{\frac{m}{2\hbar t}}\left[x-\frac{\hbar q}{m}t\right]. 
\label{eq:5}
\end{equation}
The $w$ function appears in many fields of physics and mathematics, so that it has been well studied 
and its properties have been well understood \cite{wfunction2}. Some computer programs are available 
for numerical calculation of the $w$ function \cite{wfunction3}. 

Let us now consider the following initial condition: 
\begin{equation}
\Psi(x,0)=
A\int_{-\infty}^\infty dk 
\left(\frac{e^{ikx}}{k-k_0+i\Delta}+\hbox{c.c.}\right), 
\label{eq:6}
\end{equation}
where $A=\sqrt{\Delta\left\{1+(\Delta/k_0)^2\right\}}/2\pi$ with $\Delta >0$, and ^^ ^^ c.c." 
stands for complex conjugate. An important feature of this $\Psi(x,0)$ is that it automatically 
vanishes for $x > 0$. This is immediately seen from the fact that the integrand $e^{ikx}/(k-k_0+i\Delta)$, 
which corresponds to the Lorentzian momentum distribution centered at $\hbar k_0$ with width 
$\hbar\Delta$, has a simple pole only in the lower-half of the complex $k$-plane. An explicit 
expression for $\Psi(x,0)$ can be easily obtained by the method of residues. We find 
\begin{equation}
\Psi(x,0)=\cases{4\pi A\,e^{\Delta x}\sin k_0 x & for $x<0$\cr 0 & for $x\ge 0$.}
\label{eq:7}
\end{equation}
This represents a wave packet. A measure of the packet width is $1/\Delta$. We can easily prove 
that the wave packet is normalized, i.e., $\int dx |\Psi(x,0)|^2=1$. If the above $\Psi(x,0)$ is 
multiplied by a constant $i/\sqrt{\Delta}$ and the limit $\Delta\to 0$ is taken, Eq. (\ref{eq:1}) 
is reproduced. The above $\Psi(x,0)$ is therefore a wave packet counterpart of the cut-off 
plane wave initial condition. Equation (\ref{eq:6}) thus leads us to a natural setup of the 
shutter problem with a normalized wave packet initial condition which vanishes automatically for 
$x>0$ due to the Lorentzian momentum distributions. 

Let us derive an analytical expression for $\Psi(x,t)$ under the initial condition given by 
Eq. (\ref{eq:7}). For definiteness, we shall limit our attention to the analytical expression 
only in the transmitted region $x>d$. We start from the following expression for the time 
evolved wave function for the transmitted region: 
\begin{equation}
\Psi(x,t)=\int_{-\infty}^\infty \frac{dk}{\sqrt{2\pi}}\, \phi(k)T(k)\, 
e^{ikx-i\hbar k^2 t/2m}, 
\label{eq:8}
\end{equation}
where $\phi(k)$ is the $k$-space wave function (i.e., the Fourier transform of the initial wave 
function) defined by 
\begin{equation}
\phi(k)=\int_{-\infty}^\infty \frac{dx}{\sqrt{2\pi}} e^{-ikx} \Psi(x,0)\label{eq:9a}.
\end{equation}
Equation (\ref{eq:8}) follows directly from the eigenfunction expansion of the wave function in 
the transmitted region. Substituting Eq. (\ref{eq:6}) into Eq.(\ref{eq:9a}), we have 
\begin{equation}
\phi(k)=\sqrt{2\pi}A\left(\frac{1}{k-k_0+i\Delta}-\frac{1}{k+k_0+i\Delta}\right). 
\label{eq:9}
\end{equation}
Next, we expand the transmission amplitude in terms of its complex poles and the corresponding 
residues by using a special form of the Mittag-Leffler theorem due to Cauchy \cite{cauchy}. 
It may be expanded as \cite{equivalence,yamadaun} 
\begin{equation}
T(k)=
\sum_{n=-\infty}^{\infty} \left(\frac{r_n}{k-k_n}+\frac{r_n}{k_n}\right). 
\label{eq:10}
\end{equation}
The substitution of Eqs. (\ref{eq:9}) and (\ref{eq:10}) into the right-hand side of 
Eq. (\ref{eq:8}) yields the following quantity: 
\begin{displaymath}
\left (\frac{1}{k-k_0+i\Delta}-\frac{1}{k+k_0+i\Delta}\right )
\left (\frac{1}{k-k_n}+\frac{1}{k_n}\right ), 
\end{displaymath}
which, upon expansion, gives four terms. To two of the four terms, we apply the partial fraction 
expansion 
\begin{eqnarray}
\frac{1}{k\pm k_0+i\Delta}\,\frac{1}{k-k_n}=\frac{1}{\pm k_0+k_n+i\Delta}\nonumber\\
\times \left(\frac{1}{k-k_n}-\frac{1}{k\pm k_0+i\Delta}\right) 
\end{eqnarray}
and express the resultant $k$ integrals in terms of $M$ functions. We then have 
\begin{eqnarray}
&&\Psi(x,t)=-2\pi i A\nonumber\\
&&\times\left[\sum_n\left(\frac{r_n}{k_0-k_n-i\Delta}+
\frac{r_n}{k_n}\right) M(x,k_0-i\Delta;t)\right.
\nonumber\\
&&-\sum_n\left(\frac{r_n}{-k_0-k_n-i\Delta}+\frac{r_n}{k_n}
\right)M(x,-k_0-i\Delta;t)\nonumber\\
&&\left.-\sum_n\left(\frac{r_n}{k_0+k_n+i\Delta}+\frac{r_n}{k_0-k_n-i\Delta}\right)
M(x,k_n;t)\right].\nonumber\\
\label{eq:11}
\end{eqnarray}
Due to Eq. (\ref{eq:10}), the first and the second sums over $n$ in the square brackets in 
Eq. (\ref{eq:11}) give $T(k_0-i\Delta)$ and $T(-k_0-i\Delta)$, respectively. We thus arrive at 
\begin{eqnarray}
\Psi(x,t)&=&-i\sqrt{\Delta\{1+(\Delta/k_0)^2\}}\nonumber\\
&&\times\Bigg[T(k_0-i\Delta)M(x,k_0-i\Delta;t)\nonumber\\
&&\phantom{\times\Bigg[}-T(-k_0-i\Delta)M(x,-k_0-i\Delta;t)\nonumber\\
&&\phantom{\times\Bigg[}-2k_0\sum_n\frac{r_n}{k_0^2-(k_n+i\Delta)^2}M(x,k_n;t)\Bigg].
\nonumber\\
\label{eq:12}
\end{eqnarray}
This is the analytical expression for the time evolved wave function in the transmitted region 
under the initial condition given by Eq. (\ref{eq:7}). The analytical solution under the cutoff 
plane wave initial condition, Eq. (\ref{eq:2}), can be correctly reproduced from Eq. (\ref{eq:12}) 
if we multiply Eq. (\ref{eq:12}) by a constant $i/\sqrt{\Delta}$ and then take the limit 
$\Delta\to 0$. It is also possible to derive an analytical expression for $\Psi(x,t)$ in terms of 
$M$ functions in other regions of space, although we concentrate on the transmitted region. 

To use Eq. (\ref{eq:12}), we have to find the poles $\{k_n\}$ and calculate the residues $\{r_n\}$ 
and the $w$ functions numerically. The residues may be calculated in general by using the simple 
relationship \cite{equivalence}, 
\begin{equation}
r_n=iu_n(0)u_n(d)e^{-ik_nd},
\label{eq:12a}
\end{equation}
where, as discussed in the appendix A of Ref. \cite{quantumshutter}, the resonant eigenfunctions 
$u_n(x)$ are solutions to the time-independent Schr\"odinger equation
\begin{equation}
\frac{d^2 u_n(x)}{dx^2}+\left[{k_n}^2-\frac{2m}{\hbar^2}V(x)\right] u_n(x)=0 
\label{eq:12b}
\end{equation}
satisfying the outgoing boundary conditions,
\begin{equation}
\left[\frac{d}{dx}u_n(x)\right]_{x=0} = -ik_nu_n(0);\,\,
\left[\frac{d}{dx}u_n(x)\right]_{x=d} = ik_nu_n(d),
\label{eq:12c}
\end{equation}
and the normalization condition,
\begin{equation}
\int_0^d u_n^2(x)dx + i {u_n^2(0)+u_n^2(d) \over 2k_n} =1.
\label{eq:12d}
\end{equation}
For a rectangular potential of height $V_0$ and width $d$, the resonant
eigenfunctions read,
\begin{equation}
u_n(x)= C_n [ e^{iq_nx} + D_n e^{-iq_nx}]\,\,\, (0 \leq x \leq d)
\label{eq:12e}
\end{equation}
where $q_n=[k_n^2-k_V^2]^{1/2}$, $k_V=\sqrt{2mV_0}/\hbar$, $D_n=(q_n+k_n)/(q_n-k_n)$ and $C_n$ may 
be obtained from the normalization condition given above. Alternatively, one may use the following 
explicit relationship between $k_n$ and $r_n$ to calculate $r_n$: 
\begin{equation}
r_n=\frac{4k_n^2(k_n^2-k_V^2)^{3/2}e^{-ik_n d}}{k_V^4(k_nd+2i)\sin(d\sqrt{k_n^2-k_V^2})}. 
\label{eq:15a}
\end{equation}
One can derive Eq. (\ref{eq:15a}) directly from $r_n=\exp(-ik_nd)/g'(k_n)$, where 
the prime stands for $d/dk$ and $g(k)=\exp(-ikd)/T(k)$; the exact analytical 
expression for the transmission amplitude $T(k)$ is available in the standard 
textbooks of quantum mechanics. 

Readers might have noticed that both Eqs. (\ref{eq:1}) and (\ref{eq:7}) correspond to an initially 
vanishing probability current density, i.e., $J(x,0)=0$. This is, however, not a general feature of 
the quantum shutter approach. The approach can be formulated even if $\Psi(x,0)=2i\sin k_0x=e^{ik_0x}
-e^{-ik_0x}$ in Eq. (\ref{eq:1}) is replaced by $\Psi(x,0)=ae^{ik_0x}+be^{-ik_0x}$ with arbitrary 
constants $a$ and $b$. 
For this general  plane wave initial condition, $J(x,0) \neq 0$ in general. In the same vein as above, 
we can construct a wave packet counterpart, for which $J(x,0)\ne 0$ as well. 
For this wave packet initial condition with nonzero flux, one may 
also obtain an analytical expression for the time evolved wave function in terms of $M$ functions. 
\begin{figure}[!btp]
\rotatebox{0}{\includegraphics[width=3.3in]{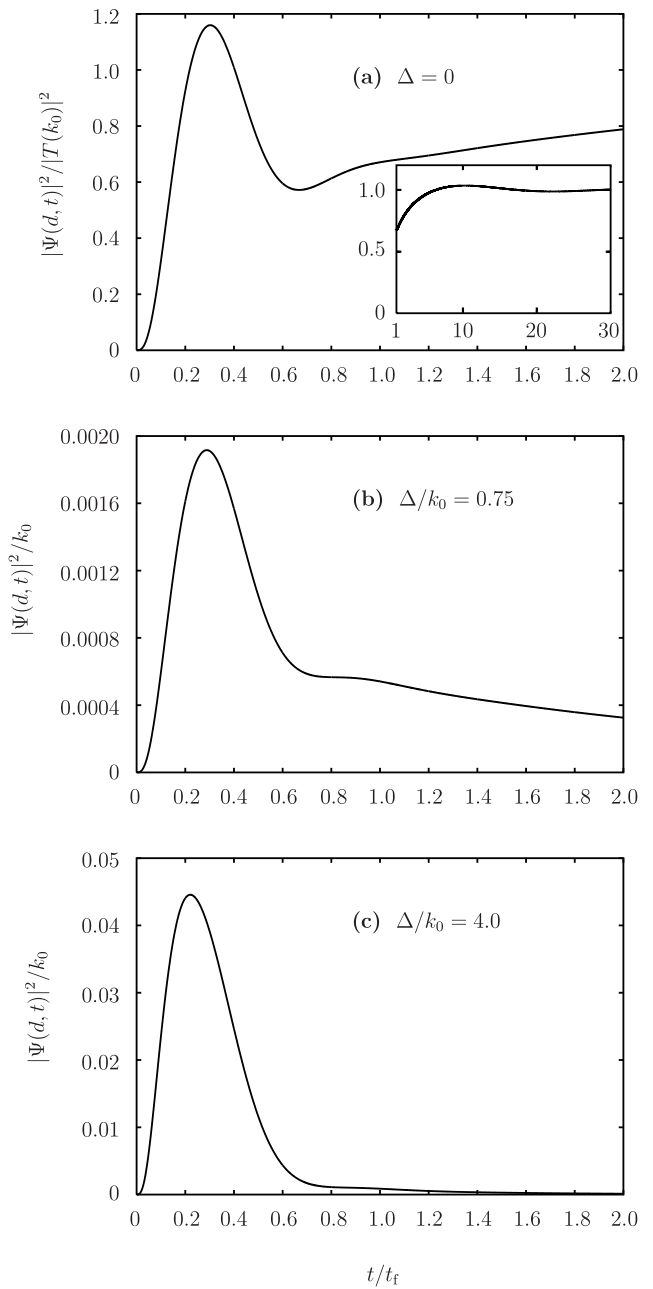}}
\caption{Plot of $|\Psi(d,t)|^2$ at the barrier edge $x=d=4$ nm as the function of time in units of 
the free passage time $t_{\rm f}$. The initial condition is given by Eq. (\ref{eq:1}) in (a) and by 
Eq. (\ref{eq:7}) in (b) and (c). We calculated numerically 1000 poles in the 3rd quadrant and also 
in the 4th quadrant in the complex $k$ plane, and used them with Eqs. (\ref{eq:3}) and (\ref{eq:12}) 
to plot these graphs.}
\label{fig1}
\end{figure}

\section{Example}

In the rest of this paper, we apply Eq. (\ref{eq:12}) to tunneling through a rectangular barrier 
to study how the time-domain resonance \cite{single1} depends on the width of the incident packets. 
Assuming a rectangular barrier of height $V_0=0.3$ eV that extends from $x=0$ to $x=4$ nm, we 
have calculated the probability density at the exit of the barrier, $|\Psi(d,t)|^2$, for a 
particle of effective mass $m=0.067m_{\rm e}$ ($m_{\rm e}$ being the bare electron mass) with 
(central) energy $E_0\equiv\hbar^2 k_0^2/2 m=0.01$ eV ($k_0\approx 0.133\ \textrm{nm}^{-1}$). 
The above refers to typical semiconductor heterostructure parameters \cite{single1}.
Figure \ref{fig1} shows the results for different values of $\Delta$, where the time axis is in 
units of the free passage time $t_{\rm f}\equiv m d/\hbar k_0\approx 17.5$ fs. 
Figure \ref{fig1} (a) shows the plot of $|\Psi(d,t)|^2$ for the case of $\Delta=0$ calculated 
from Eq. (\ref{eq:2}); we can see a transient behavior of the probability density that starts 
from zero, increases to the maximum, and then approaches its asymptotic value $|T(k_0)|^2$ as 
shown in the inset of Fig. 1 (a). The peaked structure is an example of the {\it time-domain 
resonance} \cite{single1}. In the present example, the time $t_{\rm p}$ that gives the maximum 
of $|\Psi(d,t)|^2$ is about $0.303 t_{\rm f}$. Figures \ref{fig1} (b) and (c) obtained from 
Eq. (\ref{eq:12}) show respectively the case of $\Delta/k_0=0.75\, 
(\Delta\approx 0.10\ \textrm{nm}^{-1})$ and the case of $\Delta/k_0=4.0\, 
(\Delta\approx 0.53\ \textrm{nm}^{-1})$. Unlike the case of Fig. \ref{fig1} (a), the probability 
density approaches zero for large values of $t$ in Figs. \ref{fig1} (b) and (c). This is simply 
because the particle is eventually reflected or transmitted for the wave packet initial 
condition. 

We first find from the graphs that the peaked nature of $|\Psi(d,t)|^2$ is not lost (i.e., the 
time-domain resonance is preserved) even for relatively large values of $\Delta$ (and thus for 
relatively narrow packets in configuration space). In the case of Fig. \ref{fig1} (b), the packet 
width $1/\Delta$ is more than twice the barrier width. In the case of Fig.~\ref{fig1}~(c), where 
the time domain resonance is still clear, the packet width is less than half of the barrier width. 

Secondly, we find that $t_{\rm p}$ does not depend significantly on the value of $\Delta$ in the
tunneling regime. In the case of Fig.~\ref{fig1}~(b), where the average energy of the particle, 
$\langle E\rangle=\{1+(\Delta/k_0)^2\}E_0$, is approximately $0.016$ eV, we have 
$t_{\rm p}/t_{\rm f}\approx 0.288$, so that the shift of $t_{\rm p}/t_{\rm f}$ from the case 
of Fig.~\ref{fig1}~(a) is only about $5 \%$. In the case of Fig.~\ref{fig1}~(c), where 
$\langle E\rangle\approx 0.17$ eV, we have $t_{\rm p}/t_{\rm f}\approx 0.221$, which shows as much 
as $27 \%$ shift from the case of Fig.~\ref{fig1}~(a). Although the average energy is below the 
barrier height in both cases, we must note that only the case of Fig.~\ref{fig1}~(b) can be 
associated with tunneling as explained below. 

In what follows we refer to two different approaches to distinguish between tunneling and non-tunneling 
processes. The first one involves a stationary analysis whereas the second one deals with a 
time-dependent description. For wave packet initial conditions, the first approach relies on the 
computation of both the tunneling probability $P_{\rm under}$ (i.e., the probability of under-the-barrier 
transmission), and the non-tunneling probability $P_{\rm over}$ (i.e., the probability of over-the-barrier 
transmission) defined by 
\begin{eqnarray}
P_{\rm under}=\int_0^{k_V}\!dk |T(k)|^2|\phi(k)|^2,
\label{under}\\[.5cm]
P_{\rm over}=\int_{k_V}^\infty dk|T(k)|^2|\phi(k)|^2.
\label{over}
\end{eqnarray}
In the case of Fig.~\ref{fig1}~(b), 
$P_{\rm under}\approx 0.00161$ and $P_{\rm over}\approx 0.00117$, so we affirm that under-the-barrier 
transmission slightly dominates, and in this sense, the transmission is in the tunneling regime. In the 
case of Fig.~\ref{fig1}~(c), $P_{\rm under}\approx 0.0111$ and $P_{\rm over}\approx 0.0426$, 
so that over-the-barrier transmission dominates and the process is not in the tunneling regime.
The second approach deals with a time-frequency analysis \cite{cohen}, where we introduce the local 
average frequency $\omega_{\rm av}$ (the instantaneous frequency of the wave function) and the 
instantaneous bandwidth $\sigma$ (the spread of frequencies around $\omega_{\rm av}$) as follows \cite{cohen}: 
\begin{eqnarray}
\omega_{\rm av}(t)&=&-{\rm Im}\,\left(\frac{1}{\Psi}\frac{\partial\Psi}{\partial t}\right),\label{eq:13}\\[.5cm]
\sigma(t)&=&\left\vert{\rm Re}\,\left(\frac{1}{\Psi}\frac{\partial\Psi}{\partial t}\right)\right\vert.\label{eq:14}
\end{eqnarray}
The local average frequency was used in Refs. \cite{mb2000,single4} and the local bandwidth in 
Ref. \cite{single4} in the studies of tunneling time. It is immediate from Eq. (\ref{eq:14}) that 
$\sigma=0$, i.e., the wave function has a single instantaneous frequency, when 
$\partial\vert\Psi(x,t)\vert^2/\partial t=0$, which holds at the time domain resonance peak at $t=t_{\rm p}$. 
Therefore, the time domain resonance peak is characterized by a single instantaneous frequency 
$\omega_{\rm av}$, or equivalently, by a single energy $\hbar \omega_{\rm av}$. This was exemplified in 
Ref. \cite{single4}. We have calculated $\omega_{\rm av}$ at the time domain resonance peak, namely, 
$\omega_{\rm av}(t=t_{\rm p})$, to find that $\omega_{\rm av}/\omega_{V_0}$, with $\omega_{V_0}=V_0/\hbar$, 
is $\approx$ $0.792$, $0.944$, and $1.583$ for the case of Fig.~\ref{fig1}~(a), (b), and (c), respectively. 
This implies that the time domain resonance peak is associated with tunneling for the case of 
Figs.~\ref{fig1}~(a) and (b), but it is not for the case of Fig.~\ref{fig1}~(c) since 
$\omega_{\rm av}/\omega_{V_0} > 1$. We have thus shown, using two different approaches, that 
Figs.~\ref{fig1}~(a) and (b) correspond to tunneling, but Fig.~\ref{fig1}~(c) does not. 

As shown above, $t_{\rm p}$ is not very sensitive to the width of the packets in the tunneling regime. 
This in turn justifies the use of the cutoff plane wave initial condition as far as the estimation 
of the value of $t_{\rm p}$ in the tunneling regime is concerned; $t_{\rm p}$ is expected to 
characterize the earliest tunneling response of the system \cite{single1} and thus, it would be 
of relevance for device applications. A comparison of $t_{\rm p}$ with other tunneling time scales, 
such as the delay time, the B\"uttiker traversal time, and the semiclassical time, has been given 
elsewhere \cite{single1}.

In Ref. \cite{single1}, the time-domain resonance was also studied with a cutoff pulse initial condition 
given by $\Psi(x,0)=2i\sin k_0 x$ for $-a\le x\le 0$, and zero otherwise. It was found that the 
resulting time-domain resonance structure is almost identical to the one obtained with the 
semi-infinite cutoff plane wave initial condition, i.e., Eq. (\ref{eq:1}). This is consistent 
with our result that the time-domain resonance structure is not quite sensitive to the packet 
width. On the other hand, the oscillatory behavior of the probability density in the ^^ ^^ post 
resonance" time domain (see Fig. 7 in Ref. \cite{single1}) cannot be seen in our results. This 
implies that the probability density in the ^^ ^^ post resonance" time domain is sensitive to 
the shape of the incident wave. 

\section{Summary}

In summary, we have extended the type of initial conditions used in the quantum shutter approach 
from shuttered plane waves to a certain class of wave packets. This makes it possible to study 
the ^^ ^^ size effect" of the packets on the various results that had been obtained from the quantum 
shutter approach with shuttered plane wave initial conditions. We have derived an analytical 
expression for the time evolved wave function, Eq. (\ref{eq:12}), under the wave packet initial 
condition given by Eq. (\ref{eq:7}). Focusing on the size effect on the time domain resonance, 
we exemplified that (i) the time-domain resonance structure is present even when the packet width 
$1/\Delta$ is much shorter than the barrier width, a situation that is quite different from those 
considered in the original quantum shutter approach, where the incident wave is semi infinite, 
and (ii) the time $t_{\rm p}$ at which the time resonance peak occurs is not very sensitive to the 
packet width when the transmission process is in the tunneling regime. In this sense, the time domain 
resonance is robust against the change of the initial conditions. 

\section{Acknowledgments}

N.Y. thanks the Department of Physics, UNAM for their hospitality. He thanks H. Yamamoto for 
encouragement. N.Y. carried out a part of his numerical calculations on SX7 at the Information 
Synergy Center, Tohoku University. G.G-C and J.V  acknowledge partial financial support of 
DGAPA-UNAM under grant No. IN108003.

\end{document}